\documentclass[useAMS,usenatbib]{mn2e}

\usepackage{graphicx,amssymb,amsmath}

\title[Biases in GCs parameters estimation]{Biases in the determination of dynamical 
parameters of star clusters: today and in the Gaia era}
\author[Sollima et al.]{A. Sollima$^{1}$\thanks{E-mail:
antonio.sollima@oabo.inaf.it}, H. Baumgardt$^{2}$, A. Zocchi$^{3}$, 
E. Balbinot$^{3}$, M. Gieles$^{3}$, 
\newauthor
V. H\'{e}nault-Brunet$^{3}$, A. L. Varri$^{4}$\\
$^{1}$ INAF Osservatorio Astronomico di Bologna, via Ranzani 1, Bologna, 40127,
Italy\\
$^{2}$ School of Mathematics and Physics, University of Queensland, St Lucia,
QLD 4072, Australia\\
$^{3}$ Department of Physics, University of Surrey, Guildford GU2 7XH, UK\\
$^{4}$ School of Mathematics and Maxwell Institute for Mathematical Sciences, University of Edinburgh, 
Edinburgh EH9 3JZ, UK
}
\begin{document}


\pagerange{\pageref{firstpage}--\pageref{lastpage}} \pubyear{2015}

\maketitle

\label{firstpage}

\begin{abstract}
The structural and dynamical properties of star clusters are generally derived
by means of the comparison between steady-state analytic models and the 
available observables. With the aim of studying the biases of this approach, we 
fitted different analytic models to simulated observations obtained from 
a suite of direct N-body simulations of star clusters in different stages of 
their evolution and under different levels of tidal stress to derive mass, 
mass function and degree of anisotropy. 
We find that masses
can be under/over-estimated up to 50\% depending on the degree of
relaxation reached by the cluster, the available range of
observed masses and distances of radial velocity measures from the cluster
center and the strength
of the tidal field. The mass function slope appears to be better constrainable and less
sensitive to model inadequacies unless strongly dynamically evolved clusters
and a non-optimal location of the measured luminosity function are considered.
The degree and the characteristics of the anisotropy developed in the N-body
simulations are not adequately reproduced by popular analytic models and can be detected 
only if accurate proper motions are available.
We show how to reduce the uncertainties in the mass, mass-function and
anisotropy estimation and provide predictions for the improvements expected when 
Gaia proper motions will be available in the near future.
\end{abstract}


\section{Introduction}
\label{intro_sec}

The internal kinematics of star clusters offers a wealth of information on
their present day dynamical status and provides precious
traces of the past history of evolution and interaction with the host galaxy of 
these stellar systems. In old and relatively low-mass stellar systems (like open
and globular clusters) the half-mass relaxation time is often shorter that their
ages and processes like kinetic energy equipartition, mass segregation and core
collapse can be at work and leave signatures in the phase-space distribution of 
their stars. The understanding of these processes is crucial to properly model
the dynamics of these stellar systems and to derive their global
characteristics (total mass, mass function, degree of anisotropy, etc.) from
observations performed in spatially restricted region of the cluster.

The easiest way to model the internal kinematics of star clusters is through
analytic models. Alternative approaches include Jeans modelling (Ogorodnikov, 
Nezhinskii \& Osipkov 1978; Mamon \& Bou{\'e} 2010),
Schwarzschild's orbit superposition method (Schwarzschild 1979; Vasiliev
2013) and comparison with Monte Carlo and N-body
simulations (Giersz et al. 2008, 2009, 2011; Heggie et al. 2014). 
However, the former two approaches do not ensure physically meaningful solutions
(i.e. the corresponding distribution function could be negative), while the
latter two require an enormous computational effort and have started to become feasible for modelling rich star
clusters only in recent years (e.g., see Heggie et al. 2014). 
Analytic models are generally defined by distribution functions
depending on constants of the motion, and assume the cluster is in a steady-state and 
in equilibrium with the surrounding tidal field. Because star clusters are good
approximation of collisions-dominated systems we have a relatively advanced
understanding of the distribution of their stars in phase-space from theory and
numerical simulations, and the choice of distribution function-based models is
justified. The most popular model of this
kind is the King (1966) model which proved to be quite effective in reproducing
the surface brightness profiles of many globular clusters (GCs), open
clusters and dwarf galaxies (Djorgovski 1993; McLaughlin \& van der Marel 2005;
Carballo-Bello et al. 2012; Miocchi et al. 2013). A generalization of this model 
accounting for radial anisotropy and a degree of equipartition among an arbitrary number of
mass components has been provided by Gunn \& Griffin (1979; see also Da Costa \&
Freeman 1976; Merritt
1981). The underlying assumptions of these models (i.e. the functional dependence of
the distribution function on the integrals of motion and masses), although 
relying on a physical basis, are only arbitrary guesses to model the result of 
the complex interplay among many physical processes.
Unfortunately, the validity of these assumptions is not easy to be tested with
real data since it would require the determination of complete radial profiles,
radial velocities and relative proper motions of large samples of stars with different
masses. However, ground based photometric observations are not able to resolve 
faint low-mass stars in the cores of the
most concentrated globular clusters (GCs), and Hubble Space Telescope (HST)
observations are often available only for a limited portion of the cluster
extent. Therefore, a single density profile is often calculated by mixing surface brightness
estimates (dominated by the contribution of red giant stars) in the innermost 
region with main sequence (MS) star counts in the outskirts.
For the same observational limits, information on the shape of the low-mass end of the mass 
function (MF) are available only for a few nearby GCs.
Moreover, accurate line-of-sight (LOS) velocities can be derived through
spectroscopic analyses only for the most massive red giant branch (RGB) stars.
An even more complex task is the derivation of proper motions: to date, the
typical uncertainty of the available astrometric surveys and their limiting
magnitude allowed the estimate of relative proper motions only for the brightest stars
of the closest GCs (McLaughlin et al. 2006; Anderson \& van der
Marel 2010). These data, while providing an indication of
the systemic cluster motion, are not accurate enough to be used for studies on
their internal dynamics. Only very recently proper motions with the right accuracy
derived from extensive HST surveys are becoming available (Bellini et al. 2014;
Watkins et al. 2015).   
In the absence of suitable observational data, Trenti \& van der Marel (2013) studied the effect of 
two-body relaxation on the projected
velocity dispersion of groups of stars with different masses in a set of 
N-body simulations. They
noted that the simulated clusters never reach complete kinetic energy
equipartition (a result already found previously by Giersz \& Heggie 1997 and 
confirmed on real GCs by preliminary results by Bellini et al. 2013) and 
argued that the widely used King-Michie models could be
inadequate to model these stellar systems.

\begin{figure*}
 \includegraphics[width=12cm]{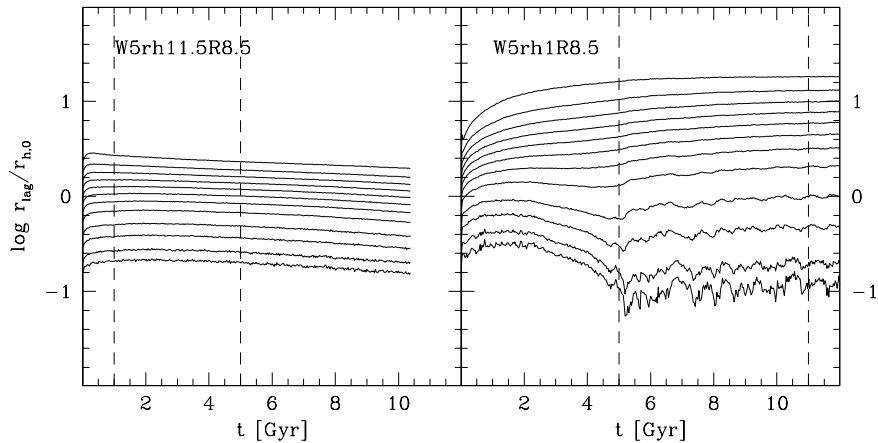}
 \caption{Lagrangian radii evolution of the N-body simulation W5rh11.5R8.5 (left
 panel) and W5rh1R8.5 (right panel). The radii containing the 1\%, 2\%, 5\%, 10\%, 20\%,
 30\%, 40\%, 50\%, 60\%, 70\%, 80\%, 90\% of the cluster mass are marked. The time at which selected snapshots are
 extracted are marked with dashed lines.}
\label{lag}
\end{figure*}

The possible inadequacy of these models could have serious consequences on the
estimate of many intrinsic properties of GCs obtained so far.
For example, since a GC is a self-gravitating system in virial equilibrium, its 
velocity dispersion can be used as a proxy for its mass. For this reason,
many studies have been devoted in the past to the determination of dynamical
masses by means of the comparison of the available photometric and kinematical observables 
(mainly projected density/surface brightness profiles and line-of-sight
radial velocities of RGB stars) with analytic models (Mandushev, Staneva \&
Spasova 1991; Pryor \& Meylan 1993; Lane et al. 2010;
Zocchi, Bertin \& Varri 2012; Kimmig et al. 2014).
Because of the above mentioned lack of information on the radial distribution
and kinematics of low-mass stars, 
these studies estimated masses through the comparison of the available data
with single-mass models (such as King 1966 models) using RGB stars as tracers of
the potential. This approximation, however, is incorrect and can lead to
significant biases in those clusters where a large number of collisions
allowed an effective exchange of kinetic energy among stars with different
masses (Shanahan \& Gieles 2015, hereafter SG15). Indeed, massive stars tend to transfer kinetic energy in collisions 
with less massive ones, becoming kinematically cold and moving on less 
energetic orbits at small radii. So, in these clusters both the spatial 
distribution and the
velocity dispersion of RGB stars significantly differ from those of the other less
massive stars which contribute to the largest fraction of the cluster mass. The
situation is further complicated by the presence of dark remnants (white dwarfs, neutron
stars and black holes) and binary stars contributing to the cluster potential
but whose fraction is hard to be estimated.
Because of these biases, the comparison of the derived dynamical masses with 
those estimated converting the cluster integrated 
luminosity into mass (often referred as 
"luminous mass") led to puzzling and conflicting results. Indeed, the first studies indicated 
that dynamical masses were $\sim$25\% smaller than luminous ones 
(McLaughlin \& van der Marel 2005; Strader et al. 2009, 2011).
The above discrepancy is partly due to the adoption of RGB stars as dynamical
tracers of the cluster potential, and to the neglecting of dynamical processes
like the preferential losses of low-mass stars altering the mass-to-light ratio
used to derive the luminous masses (Kruijssen \& Mieske 2009).
The above effects have been accounted for in a recent analysis by 
Sollima, Bellazzini \& Lee (2012) who compared deep 
HST photometric data and an extensive
survey of radial velocities for six Galactic GCs with a set of multi-mass
King-Michie models, simultaneously fitting the luminosity
function estimated in the innermost cluster region, the surface brightness and 
the velocity dispersion profile of RGB stars to derive global masses and MFs. 
Adopting the recipe of mass segregation of these models, they found
that dynamical masses are on average $\sim$40\% larger than luminous ones
i.e. in the opposite direction of what found in previous analyses.
In this case, while a large portion of this discrepancy can be due to an uncertain
assumption of the fraction of dark remnants, a significant contribution could be
due to the inadequacy of the adopted models in reproducing the degree of mass
segregation of the analysed clusters.

Another aspect connected to the modelling of mass segregation of GCs regards the
derivation of their global MF.
Indeed, thanks to the proximity of many GCs and the high angular resolution of HST 
it has been possible to sample their luminosity
function down to the hydrogen burning limit (Piotto, Cool \& King 1997; 
Piotto \& Zoccali 1999; Paust et al. 2010). These kind of observations are
however limited to a restricted portion of the cluster, generally close to the 
half-mass radius (where the effects of mass segregation are minimized; Pryor,
Smith \& McClure 1986; Vesperini \& Heggie 1997; Baumgardt \& Makino 2003) or the cluster center (see e.g. the {\it "ACS globular 
clusters treasury project"}; Sarajedini et al. 2007).
Because of mass segregation, to derive the global MF from the local
one a correction based on multi-mass modelling is necessary. 
This task has been performed by Paust et al. (2010) and Sollima et al. (2012)
who found that the MFs of their samples of GCs can be well reproduced
by power-laws with slopes varying in a wide range ($-1.7<\alpha_{MF}<-0.2$).
Any possible bias in the adopted prescription for the distribution of masses
would translate into a systematic error in the estimated global MF.

Finally, it is not clear if the velocity distribution of GCs stars is isotropic
or if it presents some radial/tangential bias. 
Physical reasons at the origin of such anisotropy can be found in their 
mechanism of formation (possibly related to the collapse of their original gas 
cloud; Lynden-Bell 1967), to the collisions occurring in their central regions
(Lynden-Bell \& Wood 1968; Spitzer \& Shull 1975) or to the interaction with the 
Milky Way (Oh \& Lin
1992). The determination of anisotropy in GCs is extremely complex since it requires
the knowledge of transverse velocities for a large sample of stars to allow an
unambiguous detection, and detailed studies have been possible only very
recently for the innermost regions of nearby GCs (Watkins et al. 2015). 
King \& Anderson (2001) suggested that the dependence of
the degree of anisotropy on mass could be different from what is predicted by
King-Michie models.

The advent of the Global Astrometric Interferometer for Astrophysics
(Gaia) will change the picture by providing accurate
distances and proper motions for hundreds of stars in a large sample of nearby 
GCs (Pancino, Bellazzini \& Marinoni 2013). Indeed, the unprecedented 
astrometric accuracy of Gaia will provide proper motions with an accuracy
comparable to that of LOS radial velocities obtainable through high resolution 
spectroscopy. These measures, complemented by
those soon available from the HST analyses in the central regions of GCs, could
open the way for a new golden age for studies on the internal dynamics of GCs.

In this paper we will compare the most widely used analytic models with a suite of
N-body simulations applying the most widely used technique to determine mass,
MF and anisotropy with the aim of studying the biases in the 
determination of these quantities linked to the
ability of analytic models to reproduce the correct phase-space structure of
the simulated stellar systems. In Sect. 2 the adopted analytic models and the
performed N-body simulations are described. In Sect. 3 we describe the method
adopted to derive the best-fit masses and MFs. In Sect. 4 the results
of the comparison between the true and the estimated masses and MFs
are presented. Sect. 5 is devoted to
the simulation of Gaia observations of GCs including realistic errors and
radial sampling efficiency to test the feasibility of studies on GCs 
anisotropy. We finally summarize our results in Sect. 6

\section{Simulations and Models}
 
\subsection{N-body simulations}
\label{nbody_sec}

The N-body simulations considered here have been performed using the collisional
N-body codes NBODY4 and NBODY6 
(Aarseth 1999) and are part of the surveys presented by Baumgardt \& Makino
(2003) and Lamers, Baumgardt \& Gieles (2013)\footnote{both simulations are
available at the "Gaia Challenge" webpage
http://astrowiki.ph.surrey.ac.uk/dokuwiki}. Each simulation contains 131072 particles with no primordial 
binaries, although a small number of binaries and triples form
through tidal capture during the cluster evolution. In this configuration, the 
total cluster mass is 71236.4 $M_{\odot}$. Particles were initially distributed 
following a King (1966) model with central dimensionless potential $W_{0}=5$,
regardless of their masses. Two simulations with different half-mass radii have 
been run (with $r_{h}=1$ and 11.5 pc, hereafter referred to as W5rh1R8.5 and
W5rh11.5R8.5, respectively).
Particle masses taken from a Kroupa (2001) MF with 
a lower mass limits of 0.1 $M_{\odot}$ and an
upper mass limit of 15
and 100 $M_{\odot}$, for the W5rh11.5R8.5 and W5rh1R8.5 simulation respectively. 
The cluster moves within a
logarithmic potential having circular velocity $v_{circ}=220~km/s$, on a
circular orbit at a distance of 8.5 kpc from the galactic center. The
corresponding initial Jacobi radius is $r_{J}=61.15$ pc, i.e. equal to the one 
of the W5rh11.5R8.5 simulation. Because of their different Roche lobe 
filling factors, the tidal field affects the two simulations in extremely 
different ways. Moreover, the initial half-mass relaxation time is significantly
longer in model W5rh11.5R8.5 ($t_{rh}=4.97$ Gyr) with respect to model
W5rh1R8.5 ($t_{rh}=0.12$ Gyr). So, while the mass of the considered 
simulations are significantly smaller than those of real GCs, they bracket the
majority of globular clusters in terms of both their half-mass relaxation time
and relative strength of the tidal field ($r_{J}/r_{h}$).
Stellar evolution has been modeled using the fitting formula of Hurley, Pols
\& Tout (2000) assuming a metallicity Z=0.001. Mass lost during stellar 
evolution is assumed to be immediately lost from the cluster. No kick velocity
has been added to stellar remnants at their birth. A certain retention fraction
of newly formed black holes and neutron stars has been assumed: in simulation W5rh11.5R8.5 no
black holes are present and all neutron stars are assumed to be retained at
their birth, while in simulation W5rh1R8.5 we assumed that 10\% neutron stars
and black holes are retained. The temperature and luminosity
of each star is recorded at each time-step according to its evolutionary stage.
Simulations have
been run until cluster dissolution, defined to be the time when 95\% of the
initial mass has been lost.
The evolution of the Lagrangian radii of the two simulations is shown in Fig.
\ref{lag}. After the initial expansion driven by stellar
evolution mass-loss, simulation W5rh1R8.5 quickly undergoes core-collapse
(after $\sim$5 Gyr), and continues its evolution expanding slowly. 
Instead, because of its longer relaxation time, simulation W5rh11.5R8.5
follows a slow evolution losing a large fraction of its stars because of the
strong tidal field.

\subsection{Analytic models}
\label{mod_sec}

Selected snapshots of the above N-body simulations have been extracted and their
projected properties (density and LOS velocity dispersion profiles) have been
compared with a set of King-Michie models (Gunn \& Griffin 1979).
These models are constructed from a lowered-Maxwellian distribution function
made by the contributions of H mass groups
\begin{eqnarray}
\label{eq_df}
f(E,L)&=&\sum_{i=1}^{H} k_{i} exp\left(-\frac{A_{i}L^{2}}{2\sigma_{K}^{2}r_{a}^{2}}\right)
\left[ exp\left(-\frac{A_{i}E}{\sigma_{K}^{2}}\right)-1 \right]\nonumber\\     
\sum_{i=1}^{H}f_{i}(r,v_{r},v_{t})&=&\sum_{i=1}^{H}k_{i}
exp\left[-A_{i}\frac{v_{t}^{2}}{2\sigma_{K}^{2}}\left(\frac{r}{r_{a}}\right)^{2}\right]\times\nonumber\\
 & &\left[exp\left(-\frac{A_{i}(v_{r}^{2}+v_{t}^{2}+2\psi)}{2\sigma_{K}^{2}}\right)-1
\right]\nonumber\\
\end{eqnarray}
where $E$ and $L$ are, respectively, the energy and angular momentum per unit mass, 
$v_{r}$ and $v_{t}$ are the radial and tangential components of the velocity, 
the effective potential $\psi$ is the difference between the cluster potential 
$\phi$ at a given radius $r$ and the potential at the cluster tidal radius 
$\psi \equiv \phi - \phi_{t}$, $A_{i}$ and $k_{i}$ are scale factors for each mass group, 
and $\sigma_{K}$ is a normalization term. Note that the above distribution
function allows for various levels of radial anisotropy but does not allow
tangential anisotropy. 
The number density and the radial and tangential components of the velocity dispersion of
each mass group are obtained by integrating Eq. \ref{eq_df}: 

\begin{eqnarray}
\label{eq_struc}
n_{i}(r)&=&4\pi
\int_{0}^{\sqrt{-2\psi}}\int_{0}^{\sqrt{-2\psi-v_{r}^{2}}} v_{t}
f_{i}(r,v_{r},v_{t}) dv_{t}dv_{r},\nonumber\\
\sigma_{r,i}^{2}(r)&=&\frac{4\pi}{m_{i} n_{i}(r)}\int_{0}^{\sqrt{-2\psi}}v_{r}^{2}\int_{0}^{\sqrt{-2\psi-v_{r}^{2}}} v_{t}
f_{i}(r,v_{r},v_{t}) dv_{t}dv_{r},\nonumber\\
\sigma_{t,i}^{2}(r)&=&\frac{4\pi}{m_{i} n_{i}(r)}\int_{0}^{\sqrt{-2\psi}}\int_{0}^{\sqrt{-2\psi-v_{r}^{2}}}
v_{t}^{3} f_{i}(r,v_{r},v_{t}) dv_{t}dv_{r}\nonumber\\
\end{eqnarray}

Equations \ref{eq_struc} can be written in terms of dimensionless quantities by substituting 
\begin{eqnarray*} 
\zeta &=&\frac{v_{r}^2}{2\sigma_{K}^{2}}, \eta =\frac{v_{t}^2}{2\sigma_{K}^{2}},\\ 
\tilde{\rho}_{i}&=&\frac{m_{i} n_{i}}{\sum_{j} m_{j} n_{0,j}}, \tilde{r}=\frac{r}{r_{c}},\\ 
W&=& -\frac{\psi }{\sigma_{K}^{2}}, \tilde{r}_{a}=\frac{r_{a}}{r_{c}},
\end{eqnarray*}
where $\rho_{0}=\sum_{j} m_{j} n_{0,j}$ is the central cluster density and 
$$r_{c}\equiv \left(\frac{9 \sigma_{K}^{2}}{4 \pi G \rho_{0}}\right)^{1/2}$$
is the core radius (King 1966).
The potential at each radius is determined by the Poisson equation
\begin{equation}
\label{eq_poiss}
\nabla^{2}\psi=4\pi G \rho
\end{equation}
Equations \ref{eq_struc} and \ref{eq_poiss} have been integrated after assuming
as a boundary condition a
value of the dimensionless potential at the center $W_{0}$ outward till the
radius $r_{t}$ at which both density and potential vanish.
The total mass of the cluster $M$ is then given by the sum of the masses of all the
groups
\begin{eqnarray*}
M&=&\sum_{i=1}^{H}m_{i}N_{i}\\
 &=&\frac{9 r_{c} \sigma_{K}^{2}}{G}\sum_{i=1}^{H}
\int_{0}^{r_{t}}\tilde{\rho}_{i}\tilde{r}^{2}d\tilde{r}\\
\end{eqnarray*}
As a last step, the above profiles have been projected onto the plane of the sky 
to obtain the surface density 
$$\Sigma_{i}(R)=2\int_{R}^{r_{t}}\frac{n_{i}r}{\sqrt{r^{2}-R^{2}}}dr$$
the velocity dispersion along the line-of-sight, and in the projected radial and
tangential directions in the plane of the sky
\begin{eqnarray}
\label{eq_proj}
\sigma_{LOS,i}^{2}(R)&=&\frac{1}{\Sigma_{i}}\int_{R}^{r_{t}}\frac{n_{i}[2\sigma_{r,i}^{2}(r^{2}-R^{2})+\sigma_{t,i}^{2}R^{2}]}{r\sqrt{r^{2}-R^{2}}}dr\nonumber\\
\sigma_{r',i}^{2}(R)&=&\frac{1}{\Sigma_{i}}\int_{R}^{r_{t}}\frac{n_{i}[2\sigma_{r,i}^{2}R^{2}+\sigma_{t,i}^{2}(r^{2}-R^{2})]}{r\sqrt{r^{2}-R^{2}}}dr\nonumber\\
\sigma_{t',i}^{2}(R)&=&\frac{1}{\Sigma_{i}}\int_{R}^{r_{t}}\frac{n_{i}\sigma_{t,i}^{2}
r}{\sqrt{r^{2}-R^{2}}}dr\nonumber\\
\end{eqnarray}
The MF of unevolved stars in an annulus defined at a given 
distance R and with a width $\Delta R$ is given by
\begin{equation}
\label{eq_mf}
N_{i}(R,\Delta R)=2\pi \mu_{i}\int_{R-\Delta R/2}^{R+\Delta R/2} R~\Sigma_{i} dR
\end{equation}
while the cumulative distribution of a sample of k bins from j to j+k
\begin{equation}
\label{eq_cum}
N(<R)=2\pi\sum_{i=j}^{j+k} \mu_{i} \int_{0}^{R} R~\Sigma_{i} dR 
\end{equation}
where $\mu_{i}=1-N_{i}^{remn}/N_{i}$ is the fraction of unevolved stars and
$N_{i}^{remn}$ is the number of remnants in the i-th mass bin.

In the above models, for any choice of the $A_{i}$ coefficients, the shape of the density and velocity dispersion profiles 
are completely determined by the parameters ($W_{0},\tilde{r}_{a},N_{i}$) while
their normalization is set by the pair of parameters ($r_{c},M$).

\begin{figure*}
 \includegraphics[width=12cm]{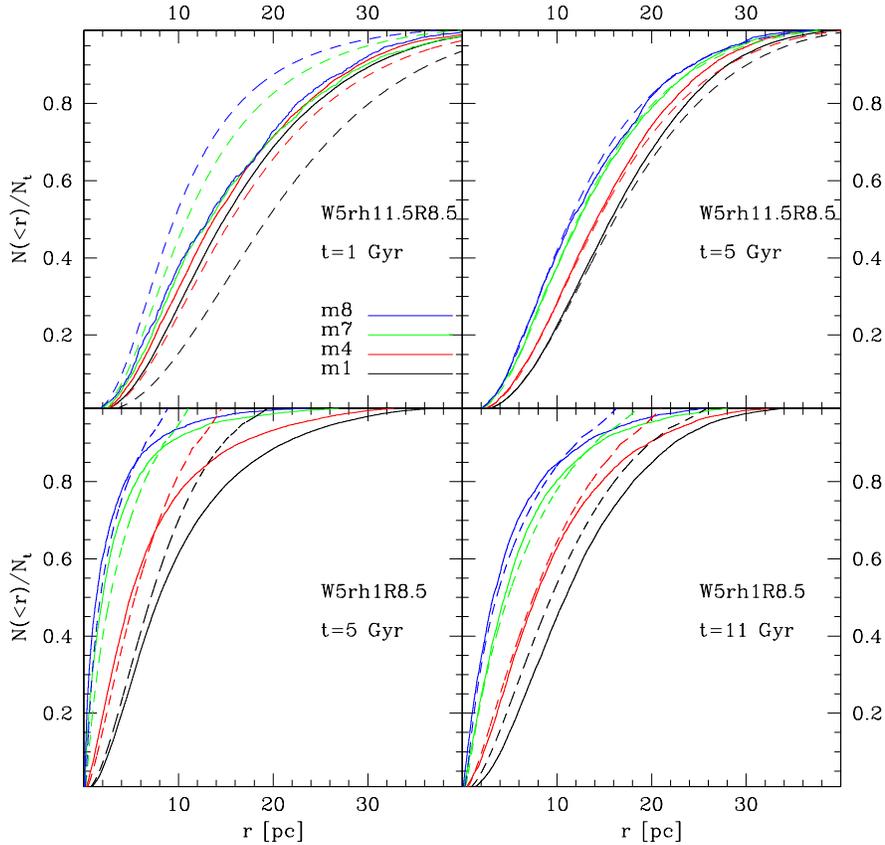}
 \caption{Cumultive radial distribution of different mass groups in the selected
 snapshots of the performed N-body simulations (solid lines). The prediction of
 the best-fit King-Michie model (assuming $\alpha=1$) are marked with dashed
 lines.}
\label{genrad}
\end{figure*}

The dependence on mass of the 
coefficients $A_{i}$ determines the degree of mass segregation of the cluster.
In the formulation by Gunn \& Griffin (1979) $A_{i}\propto m_{i}$ which implies
that more massive stars are kinematically colder than less massive ones.
Note however that, according to Eq. \ref{eq_struc} and assuming isotropy, the 
actual squared velocity dispersion of the i-th mass group is
$$\sigma_{i}^{2}=\frac{6\sigma_{K}^{2}}{5 A_{i}}\frac{\gamma(7/2,A_{i}
W)}{\gamma(5/2,A_{i} W)}$$
where $\gamma(a,b)\equiv \int_{0}^{b} x^{a-1} e^{-x} dx$ is the lower incomplete
gamma function. According to the above equation, if $A_{i}\propto m_{i}$ 
complete equipartition is
achieved only in the limit $W \rightarrow +\infty$. 
So, these models do {\it not} assume kinetic energy equipartition
neither locally nor globally (see also Merritt 1981). A slight modification to
the Gunn \& Griffin (1979) assumption can be made to allow intermediate levels
of relaxation, by simply assuming $A_{i}\propto m_{i}^{\alpha}$. In this case
$\alpha=0$ models correspond to single-mass King (1966) models, $\alpha=1$ to
Gunn \& Griffin (1979) models. We note that for $\alpha\rightarrow +\infty$ the
distribution function of low-mass stars tends asymptotically to $f(E)\propto
-A_{i} E/2 \sigma_{K}$ i.e. there is a maximum degree of kinetic
energy equipartition reachable by these models. Moreover, extrapolation at 
values of $\alpha>>1$ produces unrealistic density and velocity dispersion profiles.

\section{method}
\label{met_sec}

\begin{figure*}
 \includegraphics[width=12cm]{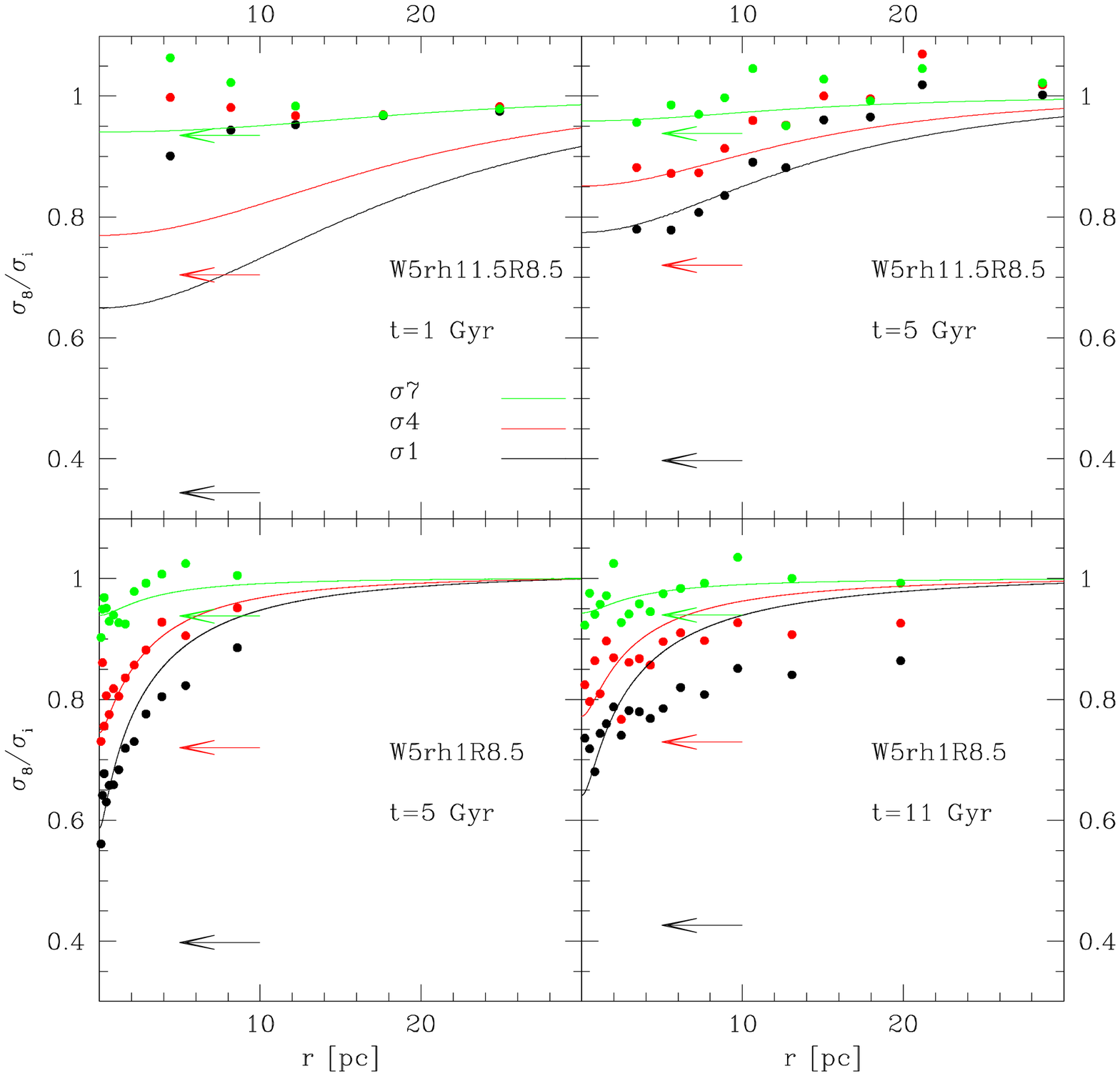}
 \caption{Ratio between the velocity dispersion of the most massive 
 and those of less massive bins as a function of distance from the cluster center for
 the selected snapshots of the performed N-body simulations.
 The prediction of the best-fit King-Michie models (assuming $\alpha=1$) are shown with solid lines.
 The values corresponding to kinetic energy equipartition are marked
 with arrows.}
\label{gensig}
\end{figure*}

The aim of this paper is to check the ability of analytic models to reproduce
the main kinematical properties of simulated star clusters. For this
purpose, we used selected snapshots of the simulations described in Sect.
\ref{nbody_sec} to produce mock observations, by considering only those
information generally available to observers. 
In particular, for each analysed snapshot we extracted:
\begin{itemize}
\item the projected number density profile of the brightest stars;
\item the MF of unevolved stars ($N_{i}^{obs}(R_{d})$) estimated in a 1 pc-width annulus around a given
distance from the cluster center (hereafter referred as the {\it deep field
range});
\item the LOS velocities of the brightest stars.
\end{itemize}
Projected distances and velocities have been calculated by projecting 3D
positions and velocities along a random direction. We considered only stars
whose projected distance lies within the Jacobi radius, to avoid the inclusion
of tidal tails in our sample.
As described in Sect. \ref{mod_sec}, analytic models are constructed from the
superposition of different mass bins. Here we considered 8 evenly spaced mass 
bins ranging from 0.1 $M_{\odot}$ to the mass at the RGB tip ($M_{tip}$), plus an additional
bin containing all the remnants more massive than $M_{tip}$. The particles of
the simulation have been divided in the above defined mass bins according to
their masses. As we are interested in systematic
biases (and not in random uncertainties) we have not added observational errors
and incompleteness effects to our data.
Moreover, we assumed that the fraction ($\mu_{i}$) and the mass distribution of dark
remnants ($N_{i}^{remn}$) are known a priori by the observer. This last assumption is far from
reality and in real cases this uncertainty can even represent the largest source of
systematics (see Sollima et al. 2012). However, as our purpose is to determine
the impact of the adopted criterion for relaxation we neglect this issue.
The impact of the uncertainties on dark remnants will be addressed in a
forthcoming paper (Peuten et al., in preparation).

The best fit to the above defined quantities has been performed using a
Markov-Chain Monte Carlo (MCMC) algorithm nested into an iterative procedure (see
Sollima et al. 2012). We performed the following steps:
\begin{enumerate}
\item We adopted as a first guess of the global MF ($N_{i}$) the sum of
the local MF of unevolved stars measured in the {\it deep field range}
($N_{i}^{obs}$) and that of remnants ($N_{i}^{remn}$);
\item the MCMC algorithm samples the parameter space
to find the pair of parameters ($W_{0},~r_{c}$) minimizing the Kolmogorov-Smirnov 
penalty function i.e. the maximum 
absolute difference between the observed and the predicted normalized cumulative 
distribution of particles in the considered mass bins (excluding remnants; 
see Eq. \ref{eq_cum}). The probability associated to the Kolmogorov-Smirnov 
test $P_{KS}$ is recorded; 
\item The MF predicted by the model in the {\it deep
field range} ($N_{i}^{mod}$) is calculated using Eq. \ref{eq_mf} and a new 
guess ($N'_{i}$) of the global MF of unevolved stars is made using the relation
$$N'_{i}=N_{i}\left(\frac{N_{i}^{obs}}{N_{i}^{mod}}\right)^{0.5}$$
where the exponent 0.5 is a damping factor to avoid divergence.
\end{enumerate}

Steps {\it (ii)} and {\it (iii)} are repeated until the maximum variation of the
estimated MF in two subsequent iterations among the various mass 
bins ($max|1-N'_{i}/N_{i}|$) falls below 1\%.

The mass of the model has been finally derived by minimizing the log-likelihood
function
\begin{equation}
\label{sig_eq}
ln L=ln P_{KS}-\frac{1}{2}\sum_{i=1}^{N}
\frac{(v_{i}-\bar{v})^{2}}{\sigma_{LOS}^{2}(R_{i})}+ln(\sigma_{LOS}(R_{i}))
\end{equation}

Where $v_{i}$ is the LOS radial velocity of the i-th star belonging to the most massive
group.
An additional cycle can be added to the above procedure to search for the 
value of the anisotropy radius $\tilde{r}_{a}$ maximizing the above defined
merit function.

\section{Results}
\label{res_sec}

\begin{figure*}
 \includegraphics[width=12cm]{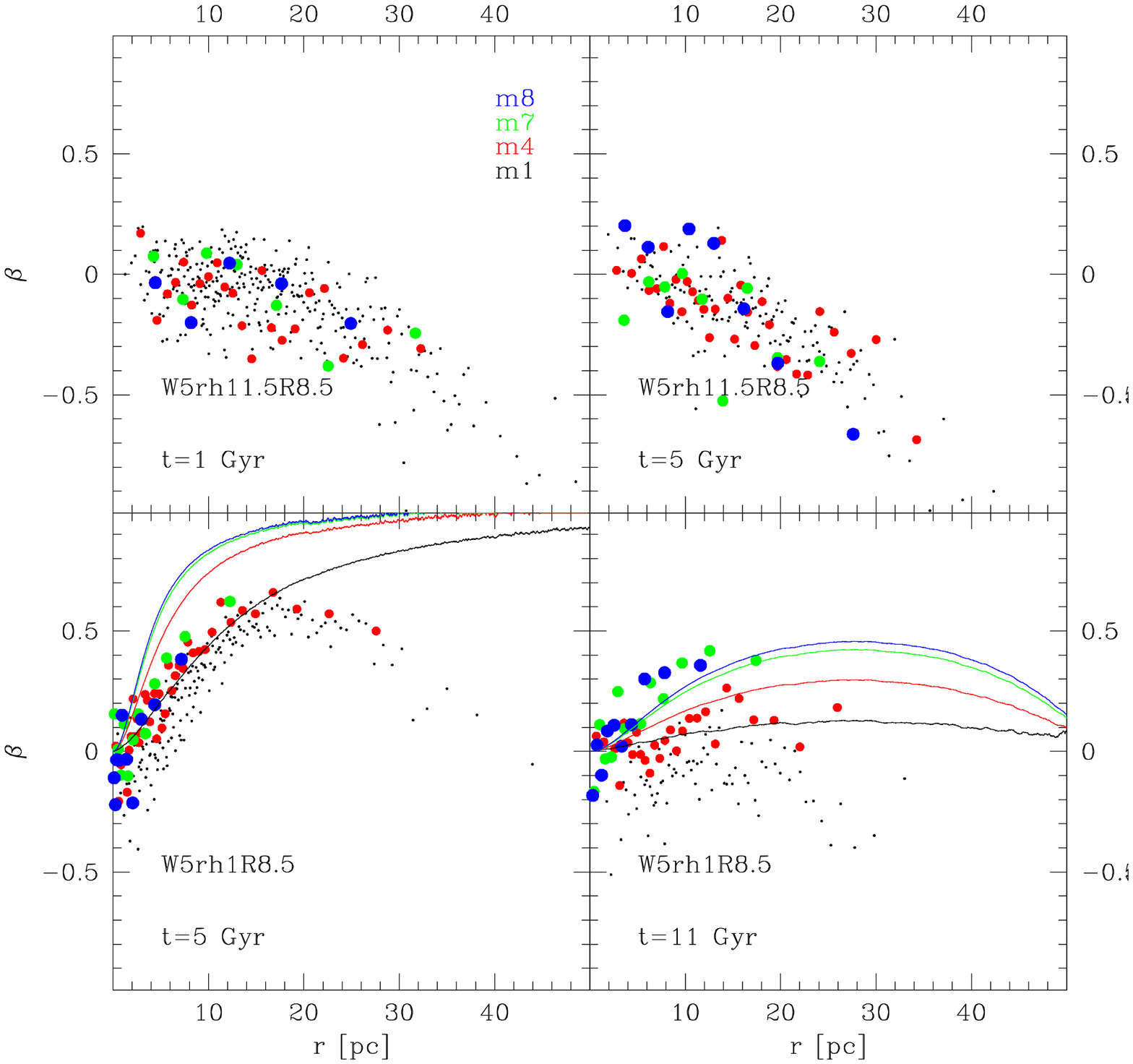}
 \caption{Anisotropy parameter as a function of the distance from the cluster
 center for various mass groups in the selected snapshots of the performed
 N-body simulations. The prediction of the best-fit King-Michie models (assuming $\alpha=1$) are marked
 with solid lines.}
\label{genanis}
\end{figure*}

As a first test we selected two snapshots for each simulation in selected
moments of the cluster evolution and compared the radial distribution, velocity
dispersion and anisotropy of different mass groups with those predicted by the
corresponding best-fit models. In particular, 
for the less evolved
simulation W5rh11.5R8.5 we selected a snapshot at 1 Gyr (after the stellar 
evolution-driven expansion and before a significant level of relaxation has 
been reached) and a snapshot at 5 Gyr ($t\sim 0.8~t_{rh}(t)$). For simulation
W5rh1R8.5 snapshots at 5 Gyr (close to core collapse) and 11 Gyr (after 
several half-mass relaxation times; $t\sim9 t_{rh}(t)$) have been selected.

In Fig. \ref{genrad} the cumulative radial distribution of four different mass
groups are compared with those predicted by the best-fit King-Michie models
assuming $\alpha$=1. As expected, in the early stage of evolution of simulation
W5rh11.5R8.5 different mass groups have still very similar distributions. 
In this case, the best-fit multi-mass model overpredicts the level 
of relaxation predicting low-mass stars distributed on a more extended structure
than they really are. On the other hand, after 5 Gyr the cluster reaches the level
of relaxation adopted by the model, which reproduces
extremely well the radial distribution of all the mass groups. A different
situation occurs in simulation W5rh1R8.5: in both the selected stages of
evolution the actual mass segregation is slightly larger than that predicted
by King-Michie models. 

The same conclusion can be reached by analysing the ratio between the 3D 
velocity dispersions\footnote{Hereafter, in the comparison between the velocity dispersion
of models and N-body snapshots the systemic motion of the cluster is subtracted 
i.e. $\sigma^{2}=\langle v^{2}\rangle-\langle v\rangle^{2}$.} of different mass bins (see Fig. \ref{gensig}). In a
relaxed stellar system kinetic energy equipartition implies that at each radius
the velocity dispersion of stars scales with their masses as 
$\sigma\propto m^{-1/2}$. So, while at the beginning of the simulation stars of
different masses have the same dispersion, as relaxation proceeds massive
stars tend to became kinematically colder than less massive ones and the ratio of their
velocity dispersions tends to the square root of their mass ratios. 
This process is faster in the center of the cluster where a large number
of collisions accelerates this process. An inspection of Fig. \ref{gensig} shows
that, as already mentioned in Sect. \ref{mod_sec}, both N-body simulations and 
analytic models are still far from complete kinetic energy equipartition even in 
the cluster center. Again, models overpredict relaxation in the first selected
snapshot of simulation W5rh11.5R8.5 while they well reproduce the velocity
dispersion ratio in all the other considered snapshots of both simulations,
except for the least massive bin which appears over-relaxed in simulation
W5rh1R8.5 than what predicted by analytical models.

\begin{figure*}
 \includegraphics[width=12cm]{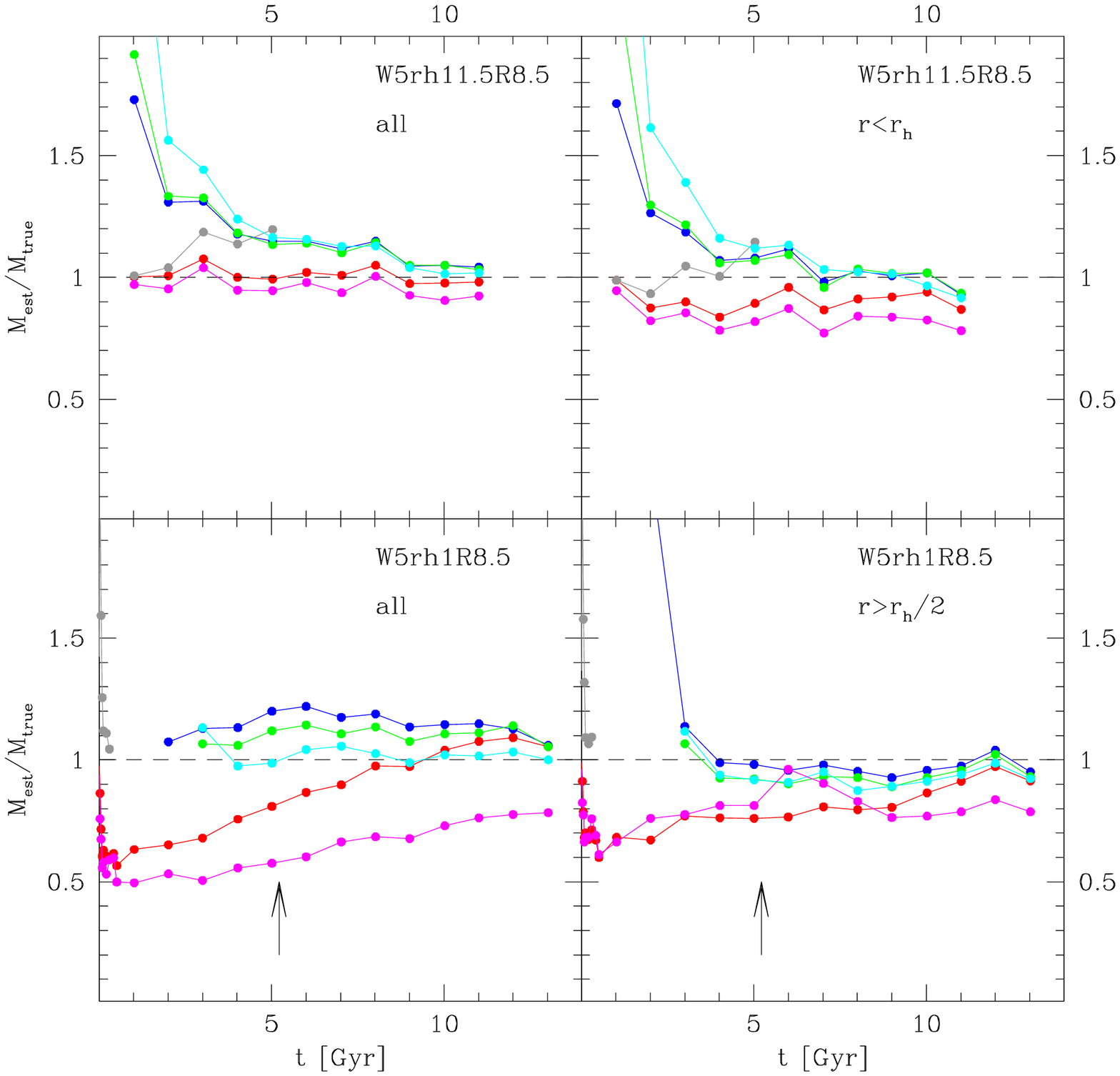}
 \caption{Ratio between the masses estimated by analytic models and the true
 cluster mass. The color code of different models in described in Sect.
 \ref{res_sec}. Left panels refer to mass estimates performed using the entire
 sample of giants, right panels refer to estimates performed applying a radial
 selection to the sample. The epoch of core-collapse in simulation W5rh1R8.5 is marked by
 an arrow in bottom panels.}
\label{mass}
\end{figure*}

In Fig. \ref{genanis} the anisotropy parameter
$\beta\equiv1-\sigma^{2}_{t}/2\sigma^{2}_{r}$ of the various mass groups is
shown as a function of the distance from the cluster center. A positive value of
$\beta$ indicates a bias toward radial orbits ({\it radial anisotropy}), while a negative value indicates
the prevalence of tangential orbits ({\it tangential anisotropy}\footnote{We
define tangential anisotropy as a bias toward tangential orbits with random
directions. Rotation, characterized by ordered motions in a single direction, is
not analysed here.}). It is apparent that the two simulations 
present opposite trends: in both the considered snapshots of simulation 
W5rh11.5R8.5 tangential
anisotropy which increases toward the outer cluster regions is apparent in all
mass bins with the same indistinguishable amplitude. Instead, in simulation
W5rh1R8.5 all mass groups quickly develop radial anisotropy. After many
relaxation times, however, this trend tends to be erased in the less-massive
stars whose orbits became again isotropic, while a significant radial anisotropy
persists in the most massive stars. The physical reason of this behaviour is
likely due to the different tidal stress experienced by the two simulations: in
simulation W5rh1R8.5 collisions occurring in the cluster center kick stars in
the cluster halo on radial orbits. This process, already studied in past studies
by Lynden-Bell \& Wood 1968 and Spitzer \& Shull 1975, appears to affect stars of
different masses with the same efficiency. In the initial stage of this
simulation the cluster is very compact and almost unaffected by the tidal field.
As evolution proceeds, the cluster expands and the kinematically hottest
low-mass stars reach the boundary of the Roche volume. In this case, stars on
radial orbits reach this boundary with positive velocity and escape more
efficiently from the cluster. This process produces the observed increasing trend of 
radial anisotropy with mass. The same effect is at work during the entire
evolution of simulation W5rh11.5R8.5: in this case, the cluster feels a
strong tidal field and stars on tangential orbits are
preferentially retained regardless of their masses.
The comparison with King-Michie models has been done only for the snapshots of
the W5rh1R8.5 simulation. Indeed, these models do not account for tangential
anisotropy present in the W5rh11.5R8.5 simulation. It is apparent that
King-Michie models predict an increasing anisotropy for more massive stars. So,
while they reproduce fairly well the snapshot at 11 Gyr,
they cannot provide a satisfactory representation of the cluster in its early
stages.

Summarizing, King-Michie models with $\alpha=1$ (i.e. $A_{i}\propto m_{i}$ as 
in the formulation by Gunn
\& Griffin 1979) appear to qualitatively reproduce the radial distribution and 
relaxation status of the performed N-body simulations after a timescale
comparable to the half-mass relaxation time. On the other hand, the
formalism of these models does not allow to reproduce the general
characteristics of the velocity anisotropy in all the stages of evolution of the simulated
clusters.

To quantify the ability of these models to reproduce the estimated mass and 
MFs we fitted models to a large number of snapshots of the two considered
simulations under different assumptions on the available range of stars used to
compute the density profile and the location of the {\it deep field} used to
estimate the local MF. In particular, we considered the following fits:
\begin{enumerate}
\item single-mass King (1966) models ($\alpha=0$) fitting the number density
profile constructed with the 4 most massive bins (red lines in Fig.s
\ref{mass} and \ref{mf});
\item single-mass King (1966) models ($\alpha=0$) fitting the number density
profile constructed with only the most massive bin (magenta lines);
\item multi-mass King-Michie models ($\alpha=1$) fitting the number density
profile constructed with the 4 most massive bins and adopting a {\it deep field
range} centered around the projected half-mass radius (blue lines);
\item multi-mass King-Michie models ($\alpha=1$) fitting the number density
profile constructed with only the most massive bin
 and adopting a {\it deep field
range} centered around the projected half-mass radius (cyan lines);
\item multi-mass King-Michie models ($\alpha=1$) fitting the number density
profile constructed with the 4 most massive bins and adopting a {\it deep field
range} centered in the cluster center (green lines);
\item multi-mass King-Michie models fitting the number density
profile constructed with the 4 most massive bins and a value of $\alpha$ chosen
to simultaneously reproduce the MF measured in two {\it deep field
ranges} centered at the projected half-mass radius and in the cluster center,
respectively (grey lines)\footnote{Because of the unrealistic behaviour of
models with large values of $\alpha$ (see \ref{mod_sec}) we performed this
comparison only with snapshots with a best-fit value of $\alpha<$1.}.
\end{enumerate}

Only isotropic models have been considered since {\it i)} as shown above, 
King-Michie models do not reproduce the qualitative trend of anisotropy
with mass, {\it ii)} they do not include a treatment for tangential anisotropy
occurring in some stage of evolution, and {\it iii)} in absence of proper motion
information the signal of radial anisotropy (characterized by an increase of the
central LOS velocity dispersion) can be mimicked by the presence of massive 
objects not (properly) accounted in the model (e.g. a large number of neutron 
stars, binaries, a sub-system of stellar mass black holes or an 
intermediate-mass black hole, etc.).
The results of the above test are described in the next subsections.

\subsection{Mass estimate}
\label{m_sec}

\begin{figure}
 \includegraphics[width=8.7cm]{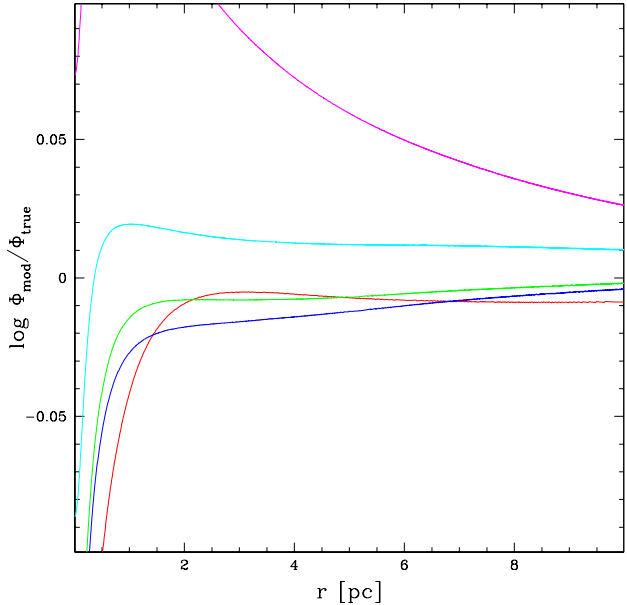}
 \caption{Ratio between the potential estimated by different models and the true
 potential of the snapshot at 11 Gyr of simulation W5rh1R8.5. The color codes of 
 different models are described in Sect.
 \ref{res_sec}.}
\label{pot}
\end{figure}

In Fig. \ref{mass} the ratio between the mass estimated by the best-fit
analytic models and the true mass of the simulation is plotted as a function
of the cluster age. By analysing simulation W5rh1R8.5 it appears that, after the 
first snapshot of the simulation, single-mass models
systematically underestimate the cluster mass up to 50\% and improve their
estimates at the end of the simulation. This is due to the quick relaxation
reached by this simulation after few Myr: as already explained in Sect.
\ref{intro_sec}, in single-mass models all stars have a single kinetic 
temperature. When a significant level of relaxation is established massive 
stars (used to sample both the number
density and the velocity dispersion of the cluster) have a smaller
velocity dispersion with respect to other cluster stars, mimicking the behaviour
of stars in the potential of a less massive cluster. Such an effect is less
evident when many mass bins are used to derive the number density profile, since
in this last case the derived mass distribution (and consequently the gravitational 
potential) is more similar to the true one. In the advanced stages of evolution
the masses estimated by single-mass models increase and exceed the real value at 
$t>$10 Gyr. This trend occurs because of the flattening of the mass function due
to the preferential escape of low-mass stars: in this stage the mean mass of
stars increases and becoming more similar to that of the stars used as tracers
of the cluster potential (see also SG15\footnote{A meaningful
comparison between these two works can be
made by comparing the $M_{est}/M_{true}$ values of the single-mass King (1966) model
fit made using only the most massive bin (magenta lines in Fig. \ref{mass}) at
ages $>$10 Gyr with the $M_{obs}/M$ plotted in the central panel of Fig. 3 in 
SG15 at a metallicity of [Fe/H]=-1 (note that their "Flat IMF" values must be
divided by 0.41 to convert $M_{SSP}$ into $M$). In both works masses
appears to be underestimated by 10-25\% (this work) and by 0-40\% (SG15) with
larger differences occurring when steeper MF are considered. Small differences
are likely due to the different stages of dynamical evolution of the clusters 
considered in the two works.}). Regarding multi-mass models with $\alpha=1$ they provide
a better estimate of the mass after $\sim$2 Gyr, although they generally
overestimate the cluster mass by 10-20\%. A counterintuitive evidence is that
models whose density profile is calculated only from the most massive bin
perform slightly better than those where more bins have been considered.
To understand the reason of such an occurrence we plot in Fig. \ref{pot} the
ratio between the potential estimated by the various models and the true one in
the snapshot at 11 Gyr. It is apparent that while all models generally
reproduce the potential at large distances from the cluster center, the central
potential is always missed by models. This region is indeed particularly 
difficult to be reproduced because massive remnants or binaries can 
sink into the very central part of the cluster producing a cusp-like shape not
reproduced by models\footnote{This effect is not removed by the adoption of 
a bin for massive remnants since it is produced by very few objects with
mass larger than the adopted bin mean mass.}. We repeated the mass estimate by 
excluding the innermost region at $r<r_{h}/2$. In this last case multi-mass
models adequately reproduce the cluster mass during the entire evolution (within
10\% except in the initial stages)
regardless of the adopted range of masses used to compute the density profile.
 
A different situation can be noted by analysing simulation W5rh11.5R8.5: here
single-mass models always well reproduce the cluster mass, while multi-mass ones
tend to overpredict the cluster mass by $\sim10\%$. Before 2 Gyr instead,
multi-mass models assuming $\alpha=1$ significantly overestimate (by more than
50\%) the mass. This is because in these early stages of evolution of this
simulation the cluster is still dynamically young and the low velocity dispersion
of massive stars predicted by models is compensated by an increased estimated
mass. The behaviour at $t>$5 Gyr is instead linked to the heating produced by
the tidal field. Indeed, because of the strong tidal field experienced by the
cluster in this
simulation, many stars reach an energy level comparable to the
potential at the tidal radius, but they take some dynamical time to escape
(Fukushige \& Heggie 2000; Baumgardt 2001). During this phase they
remain trapped within the cluster on energetic orbits at large radii with an
increased velocity dispersion. In fact, by excluding the outermost region of the
cluster (at $r>r_{h}$ where the orbits of these stars are preferentially confined) the
estimated masses decrease and bring back the estimates of multi-mass models on the
right value. It is worth noting that the best estimate is obtained constraining
the value of $\alpha$ with two measures of the MF at different
distances from the cluster center.  

\subsection{Mass Function estimate}
\label{mf_sec}

\begin{figure*}
 \includegraphics[width=12cm]{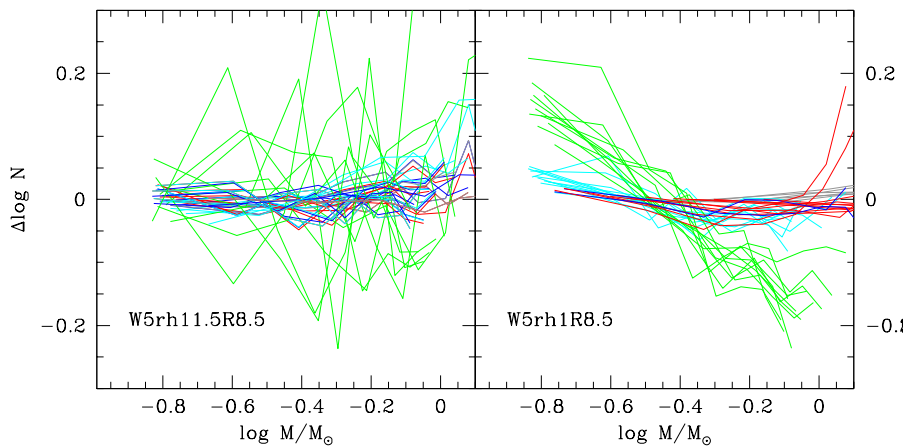}
 \caption{Difference between the actual global MFs and those
 estimated using various assumptions for the location of the {\it deep field
 range} and the mass range of stars used to compute the density profile for the
 two considered simulations. The
 color code is described in Sect. \ref{res_sec}. Red lines in both panels indicate
 the differences with respect to the MF observed at the half-mass
 radius without applying any correction.}
\label{mf}
\end{figure*}

In Fig. \ref{mf} the differences between the true and the estimated global MFs 
are shown for various assumptions and snapshots of the two considered
simulations. Of course, only multi-mass models have been considered in this
case. It is apparent that the various assumptions on the
number of bins used to construct the density profile and the choice of the
correct value of $\alpha$ do not produce significant differences in the 
determination of the MF,
providing during all the evolution a good estimate of the MF slope
(within $\Delta \alpha_{MF}<0.05$). A systematic trend is instead visible in
simulation W5rh1R8.5 when the global MF is measured by correcting
that estimated in the cluster center: in this case a slope shallower by
$\Delta \alpha_{MF}\sim0.4$ is estimated. This is a consequence of the
insufficient level of relaxation of multi-mass models which is particularly
evident in the cluster center (see Sect.
\ref{m_sec}). Consequently, the lack of low-mass stars caused by mass
segregation is erroneously interpreted as a real deficiency of stars and
therefore lead to an underestimate of the MF slope. This bias is instead
not visible in simulation W5rh11.5R8.5 (where this last approach is anyway the
most uncertain) because in this simulation multi-mass models
better reproduce the relaxation of the system.
The difference between the actual global MF and that measured at the half-mass 
radius (without applying any
correction) is also plotted in Fig.
\ref{mf} for comparison. It is apparent that also in this case, no significant
differences are noticeable.

\section{Simulation of Gaia observations}

In the previous sections we estimated the masses and the MFs of
simulated clusters through the comparison with steady-state analytic models
using as mass tracer the LOS velocity dispersion of the most massive cluster 
stars. In general, good estimates can be obtained with the desired level
of accuracy provided that radial velocities for a large enough number of stars
are available. The use of proper motions can improve the accuracy of the above 
estimates increasing the number of velocities used to constrain the cluster 
mass. However, the invaluable improvement provided by proper motions resides in
the possibility of {\it i)} testing the level of relaxation by
measuring the velocity dispersion of stars with different masses, and {\it ii)}
measuring the characteristics of anisotropy. For the first task it is necessary to measure proper
motions for MS stars down to a few magnitudes below the turn-off
point. The faint limiting magnitude and the high angular resolution needed for
this task can be achieved only with HST and will be soon available (see Bellini
et al. 2014). The second task can be instead obtained even using proper motions
for only the brightest massive stars. Gaia will provide proper motions for stars
brighter than $g<20$ over the full sky, sampling also the RGB stars of a number
of nearby GCs. An, Evans \& Deason (2012) showed that the proper motions
provided by Gaia will allow an accurate estimate of masses in GCs up to
distances of $\sim$20 kpc. In this section we
want to test if a level of anisotropy like the one developed by the simulations 
presented here can be detected with the accuracy of Gaia data. 

In principle, we could use for this test the snapshots of the performed N-body
simulations. However, the simulations considered here have masses $\sim$10 times
smaller than a typical GCs, so the velocity dispersions would be 
significantly smaller and the crowding conditions would be less severe than
in a real GC. For this reason we simulated a set mock clusters from the potential and
the phase-space distribution of the best-fit anisotropic multi-mass King-Michie 
model of the snapshot at 11 Gyr of the simulation W5rh1R8.5, but assuming
cluster masses
between $4.5<log(M/M_{\odot})<6$. Positions and velocities have been projected
and transformed in angular distances and proper motions assuming various
heliocentric distances.
Gaia magnitudes have been calculated by interpolating the particle masses 
through an isochrone by 
Marigo et al. (2008) with suitable age and metallicity. A reddening of
$E(B-V)=0.1$ and the extinction coefficients by Chen et al. (2014) have been assumed.
Errors on proper motions have been added using the software PyGaia provided by the
Gaia Project Scientist Support team\footnote{http://github.com/agabrown/PyGaia} 
as a function of magnitude,
color and assuming a K spectral type (suitable for RGB stars). These information
are computed as all sky average, neglecting efficiency variations on the sky,
and assuming the after-launch throughputs and error curves.  
Because of the angular size of the readout window, proper motions can be computed only
for relatively isolated stars. Here we consider only stars with no neighbors
brighter than 2.5 magnitudes with respect to the target g magnitude
within $3.54\arcsec$ (see Pancino et al. 2013).  
Gaia will also provide
radial velocities from low-resolution spectra. However, the estimated
velocity errors and the limiting magnitudes are of lower quality with respect to
those of available spectrographs mounted on 8m-class telescopes, so we
considered errors of 0.5 km/s on radial velocities only for stars brighter than 
$g<17$ assuming they come from a different source.

The anisotropy parameter $\beta$ is a function of the ratio between the tangential
and the radial components of the 3D velocity dispersion ($\sigma_{r}^{2}$ 
and $\sigma_{t}^{2}$, respectively). These quantities are 
related to the velocities along the LOS ($v_{LOS}$) and projected 
on the plane of the sky ($v_{\alpha cos\delta}$ and $v_{\delta}$) by Eq.
\ref{eq_proj} where
\begin{eqnarray*}
x&=&(\alpha-\alpha_{0})~cos\delta\\
y&=&\delta-\delta_{0}\\
R&=&\sqrt{x^{2}+y^{2}}\\
r&=&\sqrt{x^{2}+y^{2}+z^{2}}\\
v_{r'}&=&\frac{x v_{\alpha cos\delta}+y v_{\delta}}{R}\\
v_{t'}&=&\frac{y v_{\alpha cos\delta}-x v_{\delta}}{R}\\
\end{eqnarray*}
$\alpha_{0}$ and $\delta_{0}$ are the right ascension and declination of the
cluster center and $z$ is the distance along the LOS from the plane passing
through the cluster center. As the distance along the LOS is unknown it is not 
possible to determine
$\beta$ directly from observations. On the other hand, it is possible to combine
Eq.s \ref{eq_proj} to obtain
\begin{eqnarray*} 
\beta'(R)&=&\frac{\int_{R}^{r_{t}}\frac{n(r) \sigma_{r}^{2} \beta
r}{\sqrt{r^{2}-R^{2}}}dr}{\int_{R}^{r_{t}}\frac{n(r) \sigma_{r}^{2} r}{\sqrt{r^{2}-R^{2}}}dr}\nonumber\\
&=&1-\frac{\sigma_{t'}^{2}}{\sigma_{r'}^{2}+\sigma_{LOS}^{2}-\sigma_{t'}^{2}}\nonumber\\
\end{eqnarray*}
Since the maximum density and radial velocity dispersion in the above equation 
occurs at $(r=R,z=0)$, the above quantity is close to $\beta (R)$.

\begin{figure}
 \includegraphics[width=8.7cm]{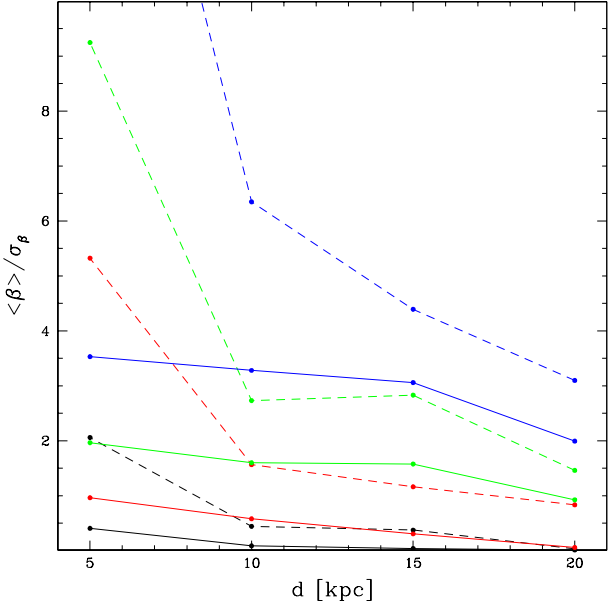}
 \caption{Significance ($\langle \beta \rangle/\sigma_{\beta}$) of the
 anisotropy detection as a function of the cluster
 distance. Black, red, green and blue lines indicate simulations of GCs with 
 masses $log(M/M_{\odot})$=4.5, 5, 5.5 and 6, respectively. Solid lines indicate
 estimates done using only stars with all the three velocity components, dashed
 lines indicate estimates done using all stars with available proper motions.}
\label{gaia}
\end{figure}

One thousand synthetic observations have been simulated for each choice of the
cluster mass and distance and the mean value of $\langle \beta' \rangle$ and its
standard deviation $\sigma_{\beta'}$ have been calculated for those stars with
both radial velocities and proper motions estimates. The $\beta'$ value of each
synthetic observation has
been calculated using the maximum-likelihood algorithm described by Eq. \ref{sig_eq}
to calculate $\sigma_{LOS},~\sigma_{r'}$ and $\sigma_{t'}$ where individual
errors on velocities ($\epsilon_{i}$) have been added to the intrinsic velocity
dispersion ($\sigma_{intr}$) assuming
$\sigma_{i}^{2}=\sigma_{intr}^{2}+\epsilon_{i}^{2}$.
The significance level (defined as
$\langle \beta' \rangle/\sigma_{\beta'}$) is plotted in Fig. \ref{gaia} as a 
function of the adopted cluster distance for different simulated masses. It can
be seen that for low/intermediate-mass clusters ($log (M/M_{\odot})<5.5$) it is
not possible to detect with a good level of significance ($>3$) the anisotropy, while a
better signal is detectable for massive ($log (M/M_{\odot})\sim6$) clusters up 
to distances of $\sim$15 kpc.

In the above test we used only stars brighter than $g<17$ with both proper
motions and LOS velocities. These two observables are however derived from
completely different techniques and are subject to different sources of
uncertainties. For instance, proper motions are measured in terms of angular
motions and require an estimate of the cluster distance to be converted into
physical units, at odds with radial velocities coming from the observed 
Doppler-shift of spectroscopic lines. Moreover, because of the different
techniques, the estimated velocity errors 
(essential to properly calculate intrinsic velocity dispersions) can present
different systematic biases. So, we considered also the case where only proper motions
are available. In this case we considered the quantity
$$\beta''=1-\frac{\sigma_{t'}^{2}}{\sigma_{r'}^{2}}$$
Although the part of the information coming from radial velocities is lost in
the above parameter ($\beta''$ is a worse estimator of $\beta$ than $\beta'$),
its corresponding uncertainty is smaller since {\it i)} it is given by
the propagation of a smaller number of uncertainties, and {\it ii)} many more 
stars can be used to compute $\beta''$ (down to $g<20$).
The significance level of this approach for the different adopted masses and
distances is overplotted in Fig. \ref{gaia}. It is apparent that this
last approach provides a larger significance level: although at large
heliocentric distances significant 
detections remain restricted to the most massive GCs, at distances $<$6 kpc it 
seems possible to detect signatures of anisotropy also in less massive clusters
($log (M/M_{\odot})>5$).

It has been shown (see Sect. \ref{res_sec}) that the level of anisotropy changes
with radius both in models and in N-body simulations. So, the values of $\beta'$ 
and $\beta''$, which are averages over stars sampled at different distance from 
the cluster center, cannot provide a fair indication of the characteristics 
(strength, radial profile, etc.) of the cluster anisotropy. A lager number of
proper motions covering the entire cluster extent are needed for this purpose.

\section{Summary}

The comparisons presented in Sect. \ref{res_sec} show that multi-mass models can
qualitatively reproduce the radial distribution and velocity dispersion of stars
with different masses in GCs subject to frequent collisions during a wide range
of their dynamical evolution. In particular, the formulation by Gunn \& Griffin
(1979; $\alpha$=1) is able to provide a fair representation of N-body
simulations after a timescale comparable to the half-mass relaxation time. By
looking at Milky Way GCs this condition is satisfied by all but three GCs
(namely $\omega$ Cen, Pal 14 and NGC2419; see McLaughlin \& van der Marel 2005;
Marin-Franch et al. 2009). During the initial half-mass relaxation time
multi-mass models overestimate the status of relaxation predicting an
unrealistic level of mass segregation. In this case, a simple generalization of
the Gunn \& Griffin (1979) formalism (adopting $A_{i}\propto m_{i}^{\alpha}$) 
provides models accounting for
intermediate levels of relaxation which better reproduce the simulation
properties. 
As a note of caution, consider that the N-body simulations used here start without any degree of
primordial mass segregation which appears to be present in many young massive
clusters (de Grijs et al. 2002; Frank, Grebel \& K{\"u}pper 2014).
After many relaxation times the analysed simulations undergo strong mass
segregation which is not reached by multi-mass models which slightly underpredict
the mass segregation present in the cluster. Both the considered simulations
quickly develop two opposite velocity anisotropy trends which cannot be reproduced by
analytic models. In the Gunn \& Griffin (1979) formalism only radially
anisotropic velocity distributions
can be simulated in King-Michie models, but alternative formulations exist to
account for tangentially anisotropic ones (see e.g. Weinberg 1991; An \& Evans
2006). However, the simple
dependence of the King-Michie distribution function from angular momentum per
unit mass cannot reproduce the wide range of trends of the anisotropy parameter
$\beta$ with stellar masses observed in N-body simulations. New models need to
be developed to model this quantity in real GCs.

Besides the above considerations, we tested the accuracy of simple fits of the
generally available observables (projected number density profiles from RGB+bright MS
stars, LOS velocities of RGB stars, MF estimated in a spatially restriced region) with 
analytic models in the estimate of mass and MF. It appears that the
adoption of single-mass models to model GCs after many half-mass relaxation
times (as widely done in many past studies because of
the lack of information on the cluster MF) can lead to underestimates of their
masses up to $\sim$50\%. This is likely the reason why dynamical masses
estimated in early studies resulted smaller than luminous ones (Mieske et 
al. 2008). The same conclusion has been reached by 
SG15 who quantified the bias in the mass estimated from 
integrated properties of a set of multi-mass GCs models if mass segregation is 
neglected and found that this effect can lead to a mass underestimate ranging
from $0.25<M_{obs}/M_{true}<1$ depending on the metallicity, the MF and the
retention fraction of dark remnants (see their Fig. 3). In the same way, the use of multi-mass
King-Michie models (with $\alpha=1$) in dynamically young GCs produces the
opposite bias leading to a significant overestimate of the cluster mass.
In this case, the adoption of models accounting for an intermediate level of
relaxation ($0<\alpha<1$) provides better results, although it requires the measure
of the MF in two different radial ranges.
Both the central cluster region and the outskirts are regions sensitive to
the uncertain distribution of heavy objects (like
neutron stars, black holes and binaries) and tidal heating. 
In this respect, tidal heating can be effective in increasing the velocity
dispersion even close to the half-mass radius and therefore the estimated mass (as
already found by K{\"u}pper et al. 2010).
An analysis devoted
to the derivation of GCs masses should therefore be restricted to a relatively
narrow portion of the cluster ($r_{h}/2<r<r_{h}$) to minimize the impact of the
above mentioned effects. In the simulations considered here, the adoption of
different samples of stars and the location of the {\it deep field} used to
measure the MF do not significantly affect the mass estimation if multi-mass
models are considered. This is due to the relatively good representation of the
mass segregation process provided by multi-mass models.

The estimate of the MF is instead less affected by the uncertainties in the
density profile and the level of relaxation. However, in
strongly mass segregated clusters the MF slope can be
underestimated by $\Delta \alpha_{MF}\sim0.4$ if the local
MF is measured at the cluster center. This is due to the inadequacy of
multi-mass King-Michie models in reproducing the relative distribution of
low/high mass stars in the central region of such relaxed GCs. Note that both
the most extensive works on the estimation of the MF slopes (Paust et al. 2010;
Sollima et al. 2012) are based on MF estimated from the same dataset of HST 
observations in the central region of GCs being potentially affected by this 
bias. This effect is however negligible in those dynamically young clusters
where multi-mass King-Michie models do a good job in reproducing the radial
distribution of different masses. It is not clear how many GCs of the above
mentioned studies can suffer from this bias.
On the other hand, the MF measured at the projected half-mass radius (without
the application of any correction) is already a fairly good representation of
the cluster global MF (see also Pryor et al. 1986, Vesperini \& Heggie 1997;
Baumgardt \& Makino 2003). So, studies based on MF 
estimated at the half-mass radius (like
those by Piotto et al. 1997 and Piotto \& Zoccali 1999) can provide good
estimates of the global MF, provided that a good estimate of the half-mass radius
is available.

A limitation of the approach adopted here is that the clusters simulated in our 
N-body simulations are a factor of 5-10 smaller than real GCs. Since the
effects of stellar evolution, relaxation and tidal effects on the dynamical
evolution of the cluster act on different timescales and their
efficiencies depend on the cluster mass in different ways, it is not possible to 
adopt some sort of scaling to GC-size objects (see Baumgardt 2001). We recall
that the simulations analysed here consider two extreme situations: while simulation 
W5rh1R8.5 is a system undergoing quick core-collapse, simulation W5rh11.5R8.5 is dynamically
young and subject to a relatively strong tidal field during all its evolution.
Thus, we expect that these simulations bracket the whole sample of possible
evolutions occurring in GCs. Anyway, any comparison with N-body 
simulations cannot be used to draw general conclusions on the dynamical
properties of real GCs. Further studies on real GCs stars are needed to assess 
the validity of the results presented here.

We explored the possibility to detect the presence and the characteristics of 
anisotropy in GCs using the proper motions provided by Gaia. We find that
moderate levels of anisotropy (like those naturally developed by N-body 
simulations) can be detected in massive ($log(M/M_{\odot})\sim6$) GCs at
distances $<$20 kpc and in intermediate-mass GCs ($log(M/M_{\odot})>5$) GCs at
$d<6$ kpc. By looking at the Harris catalog (Harris 1996; 2010 
edition) there are 25 GCs in this radial/mass range. 
Unfortunately, with the information provided by
Gaia only, it will not be possible to sample the number of objects needed to construct 
anisotropy profiles even for the
nearest GCs. Indeed, the limiting magnitude and the spatial resolution of Gaia 
will allow to obtain proper motions with errors better than 
$\sigma_{v}<2~km/s$ only for RGB stars in the outskirts of GCs. Nevertheless, 
Gaia data will be complemented by HST proper motions of stars in the
central regions of GCs thus providing the needed coverage for these kind of
studies.

\section*{Acknowledgments}

We thank the anonymous referee for his/her helpful comments and suggestions.
AS acknowledges the PRIN INAF 2011 "Multiple populations in globular
clusters: their role in the Galaxy assembly" (PI E. Carretta). 
We warmly thank Michele Bellazzini for useful discussions.
We used Gaia Challenge data products available at
{\it http://astrowiki.ph.surrey.ac.uk/dokuwiki}.

\label{lastpage}

\end{document}